\documentclass[preprint,12pt,authoryear]{elsarticle}

\usepackage{amssymb}
\usepackage{amsthm}
\usepackage{amsbsy}
\usepackage{amsmath}
\usepackage{graphicx}
\usepackage{booktabs,natbib,url}
\usepackage{subfig}

\def\diag{\mathop{\rm diag}\nolimits}
\def\tr{\mathop{\rm tr}\nolimits}

\newcommand{\ba}{\boldsymbol{a}}
\newcommand{\bb}{\boldsymbol{b}}

\newcommand{\bx}{\boldsymbol{x}}
\newcommand{\by}{\boldsymbol{y}}

\newcommand{\bA}{\boldsymbol{A}}
\newcommand{\bB}{\boldsymbol{B}}

\newcommand{\bD}{\boldsymbol{D}}
\newcommand{\bE}{\boldsymbol{E}}

\newcommand{\bI}{\boldsymbol{I}}

\newcommand{\bS}{\boldsymbol{S}}
\newcommand{\bT}{\boldsymbol{T}}
\newcommand{\bU}{\boldsymbol{U}}
\newcommand{\bV}{\boldsymbol{V}}

\newcommand{\bY}{\boldsymbol{Y}}

\newcommand{\bmu}{\boldsymbol{\mu}}

\newcommand{\bSigma}{\boldsymbol{\Sigma}}

\newcommand{\Real}{\mathbb{R}}

\journal{}

\begin{document}

\begin{frontmatter}



\title{Stable Estimation of a Covariance Matrix Guided by Nuclear Norm Penalties}


\author[label1]{Eric C. Chi}
\author[label2]{Kenneth Lange}
\address[label1]{Department of Electrical and Computer Engineering, Rice University, Texas, USA}
\address[label2]{Departments of Human Genetics, Biomathematics, and Statistics, University of California, Los Angeles, California, USA}

\begin{abstract}
Estimation of covariance matrices or their inverses  plays a central role in many statistical methods.
For these methods to work reliably, estimated matrices must not only be invertible but also well-conditioned.
In this paper we present an intuitive prior that shrinks the classic sample covariance estimator towards a stable target.
We prove that our estimator is consistent and asymptotically efficient. Thus, it gracefully transitions towards the sample covariance
matrix as the number of samples grows relative to the number of covariates. We also demonstrate the utility of our estimator in two standard situations -- discriminant analysis and EM clustering -- when the number of 
samples is dominated by or comparable to the number of covariates.
\end{abstract}


\end{frontmatter}

\section{Introduction}
\label{sec:introduction}

Estimates of covariance matrices and their inverses play a central role in many core statistical methods, ranging from least squares regression to EM clustering. In these applications it is crucial to obtain estimates that are not just non-singular but also stable. It is well known that the sample covariance matrix
\begin{eqnarray*}
\bS & = & {1 \over n} \sum_{j=1}^n (\by_j-\bar{\by})(\by_j-\bar{\by})^t
\end{eqnarray*}
is the maximum likelihood estimates of the  
population covariance $\bSigma$ of a random sample $\by_1,\ldots,\by_n$ 
from a multivariate normal distribution.  When the number of components $p$
of $\by$ exceeds the sample size $n$, the sample covariance $\bS$ is no longer invertible.
Even when $p$ is close to $n$, $\bS$ becomes unstable in the sense that small perturbations in measurements can lead to disproportionately large fluctuations in its entries. A reliable way to combat instability is to perform penalized maximum likelihood estimation.

\begin{figure}
\includegraphics[scale=0.55]{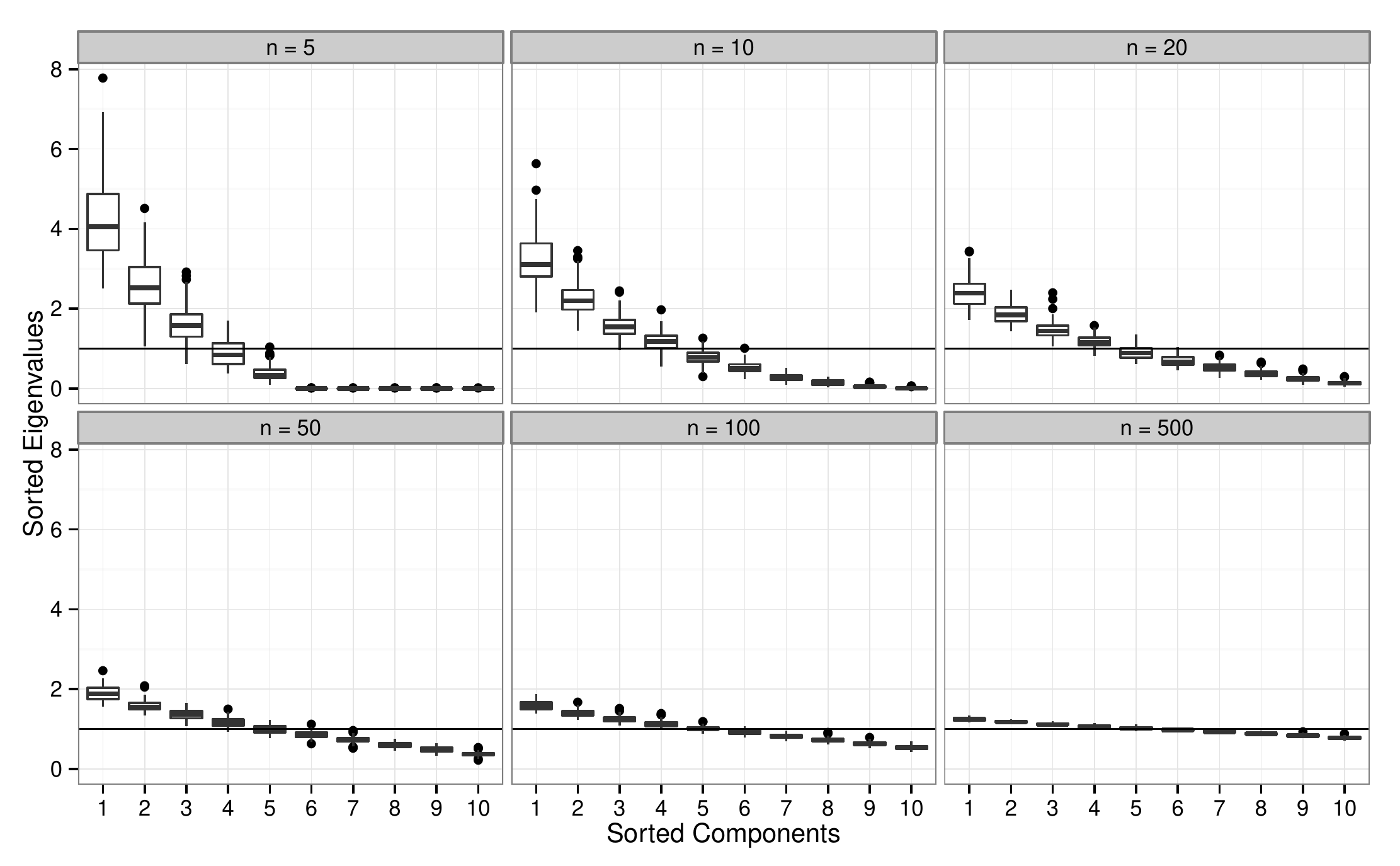}
\caption{Boxplots of the sorted eigenvalues of the sample covariance matrix $\bS$ over 100 random trials. Here the number of components $p=10$, and the sample size $n$ is drawn from the set $\{5, 10, 20, 50, 100, 500\}$.}
\label{fig:eigen_disperse}
\end{figure}

To motivate our choice of penalization, consider the eigenvalues of the sample covariance
matrix in a simple simulation experiment. We drew $n$ independent samples from a 10-dimensional multivariate normal distribution $\by_i \sim N({\bf 0}, \bI_{10})$. Figure~\ref{fig:eigen_disperse} presents boxplots of the sorted eigenvalues of the sample covariance matrix $\bS$ over 100 trials for sample sizes $n$ drawn from the set $\{5, 10, 20, 50, 100, 500\}$. The boxplots descend from the largest eigenvalue on the left to the smallest eigenvalue on the right. The figure vividly illustrates the previous observation that the highest eigenvalues tend to be inflated upwards above 1, while the lowest eigenvalues are deflated downwards below 1 \citep{LedWol2004, LedWol2012}. In general, if the sample size $n$ and the number of components $p$ approach
$\infty$ in such a way that the ratio $p/n$ approaches $\zeta \in (0,1)$, then the eigenvalues of $\bS$
tend to the Mar\^{c}enko-Pastur law \citep{MarPas1967}, which is supported on the interval $([1-\sqrt{\zeta}\,]^2, [1+\sqrt{\zeta}\,]^2)$. Thus, the distortion worsens as $\zeta$ approaches $1$.
The obvious remedy is to pull the highest eigenvalues down and push the lowest eigenvalues up. 

In this paper, we introduce a novel prior which effects the desired adjustment on the sample eigenvalues. 
Maximum a posteriori (MAP) estimation under the prior boils down to a simple nonlinear transformation of the sample eigenvalues. In addition to proving that our estimator has desirable theoretical properties, we also demonstrate its utility in extending two fundamental statistical methods -- discriminant analysis and EM clustering - to contexts where the number of samples $n$ is either on the order of or dominated by the number of parameters $p$.

The rest of our paper is organized as follows. Section \ref{sec:related_work} discusses the history of robust estimation of structured and unstructured covariance matrices. Section \ref{sec:mainresults} specifies our Bayesian prior and derives the MAP estimator under the prior. Section \ref{sec:asymptotics} proves that the estimator is consistent and asymptotically efficient. Section \ref{sec:comparisons} reports finite sample studies comparing our MAP estimator to relevant existing estimators. Section \ref{sec:applications} illustrates the estimator for some common tasks in statistics. Finally, Section \ref{sec:discussion} discusses limitations, generalizations, and further applications of the estimator. 

\section{Related Work \label{sec:related_work}}

Structured estimation of covariance matrices can be attacked from two complementary perspectives: generalized linear models and regularization \citep{Pou2011,Pou2013}. In this work we consider the problem from the latter perspective.
Regularized estimation of covariance matrices and their inverses has been a topic of intense scrutiny 
\citep{WuPou2003,BicLev2008}, and the current literature reflects a wide spectrum of structural assumptions. For instance, banded covariance matrices make sense for time series and spatial data, where the order of the components is important. It is also helpful to impose sparsity on a covariance matrix, its inverse, or its factors in a Cholesky decomposition or other factorization \citep{HuaLiuPou2006,RohTsy2011,CaiZho2012,RavWaiRas2011,RajMasCar2008,KhaRaj2011,FanLiaMin2011,BanEl2008,FriHasTib2008,HerRaj2011,HerRaj2012,PenWanZho2009}.

In this current paper, we do not assume any special structure. Our sole concern is to directly address the distortion in the eigenvalues of the sample covariance matrix. Thus, we work in the context of rotationally-invariant estimators first proposed by \cite{Ste1975}.  If $\bS = \bU \bD \bU^t$ is the spectral decomposition of $\bS$, then Stein suggests alternative estimators of the form 
\begin{eqnarray*}
\hat{\bSigma} & = & \bU \diag(e_1, \ldots, e_p)\, \bU^t
\end{eqnarray*}
that change the eigenvalues but not the eigenvectors of $\bS$. In particular, \cite{Ste1975, Haf1991, LedWol2004} and \cite{War2008} study the family 
\begin{eqnarray}
\hat{\bSigma} & = & (1-\gamma) \bS + \gamma \bT \label{warton_est}
\end{eqnarray}
of linear shrinkage estimators, where $\gamma \in [0,1]$ and $\bT = \rho \bI$ for some $\rho > 0$.
The
estimator (\ref{warton_est}) obviously entails
\begin{eqnarray*}
e_i & = & (1-\gamma)d_i + \gamma \rho .
\end{eqnarray*}
A natural choice of $\rho$, and one taken by the popular estimator of \cite{LedWol2004}, is the mean of the sample eigenvalues, namely $\hat{\sigma} = (1/p)\tr(\bS)$.  Under the assumption that $\by_i \sim N({\bf 0}, \sigma \bI)$, $\hat{\sigma}$ is the maximum likelihood estimate of $\sigma$. With this choice, the linear estimator becomes a mixture of the covariance model with the greatest number of degrees of freedom and the simplest non-trivial model. For the rest of this paper, we will assume that $\rho = \hat{\sigma}$. We also highlight the fact that the Ledoit and Wolf linear estimator, which we refer by acronym LW, is a notable member of the class of linear estimators as it specifies an asymptotically optimal value for $\gamma$ based on the data.

\cite{LedWol2004,LedWol2012} show that linear shrinkage works well when $p/n$ is large or the population eigenvalues are close to one another. On the other hand, if $p/n$ is small or the population eigenvalues are dispersed, linear shrinkage yields marginal improvements over the sample covariance. Nonlinear shrinkage estimators may present avenues for further improvement \citep{DeySri1985,DanKas2001,SheGup2003,PouDanPar2007,LedWol2012,WonLimKim2012}. Our shrinkage estimator is closest in spirit to the estimator of \cite{WonLimKim2012}, who put a prior on the condition number of the covariance matrix. 

Recall that the condition number $\kappa$ of a matrix is the ratio of its largest singular value to its smallest singular value. For a symmetric matrix, the singular values are the absolute values of the eigenvalues, and for a
covariance matrix they are the eigenvalues themselves. The best conditioned matrices are multiples of the identity matrix and have $\kappa = 1$. A well-conditioned covariance estimate is one where $\kappa$ is not too large, say in excess of 1000. 

When $n$ does not greatly exceed $p$, Figure~\ref{fig:eigen_disperse} shows that the sample covariance matrix
is often ill conditioned.  To address this defect, Won et al.\@ perform maximum likelihood estimation restricted to the space of positive definite matrices whose condition number does not exceed a threshold $\kappa_{\max}$. Let $\ell(\bSigma)$ denote the negative loglikelihood, namely
\begin{eqnarray*}
\ell(\bSigma) & = &  {n \over 2}\ln \det \bSigma+{n \over 2}\tr(\bS \bSigma^{-1}).
\end{eqnarray*}
Won et al.\@ seek an $\bSigma$ that solves
\begin{eqnarray*}
& \text{minimize} & \ell(\bSigma) \\
& \text{subject to} & \lambda_{\max}(\bSigma) / \lambda_{\min}(\bSigma) \leq \kappa_{\max},
\end{eqnarray*}
where $\lambda_{\max}$ and $\lambda_{\min}$ are the largest and smallest eigenvalues of $\bSigma$ respectively and $\kappa_{\max} \geq 1$ is a tuning parameter. Note that when $\kappa_{\max} = 1$, the unique solution is $\hat{\sigma}\bI$. In practice, $\kappa_{\max}$ is determined by cross-validation.
We will refer to the solution to the above problem as Won et al.'s CNR for condition number regularized estimator.

Won et al.\@ show that CNR has improved finite sample performance compared to linear estimators in simulations, but the greatest gains arise when only a few eigenvalues of the population covariance are much larger than the rest. The gains diminish when this does not hold. The main contribution of the estimator we describe next is 
that it provides superior performance in scenarios where CNR loses its competitive edge.

\section{Maximum a Posteriori Estimation with a Novel Prior}
\label{sec:mainresults}

Adding a penalty is equivalent to imposing a prior $\pi(\bSigma)$
on the population covariance $\bSigma$. The prior we advocate is designed to steer the eigenvalues of $\bSigma$
away from the extremes of 0 and $\infty$.  Recall that the nuclear norm of a matrix $\bSigma$, which we denote by $\| \bSigma \|_{*}$, is the sum of 
the singular values of $\bSigma$. The reasonable choice 
\begin{eqnarray*}
\pi(\bSigma) & \propto & e^{- {\lambda \over 2}\left[\alpha \|\bSigma\|_{*}+(1-\alpha)\|\bSigma^{-1}\|_{*}\right]},
\end{eqnarray*}
relies on the nuclear norms of $\bSigma$ and $\bSigma^{-1}$, a positive
strength constant $\lambda$, and an mixture constant $\alpha \in (0,1)$.

This is a proper prior on the set of invertible matrices. One can demonstrate this fact by comparing the nuclear norm $\| \bSigma \|_{*}$ to the Frobenius norm $\|\bSigma\|_{\mathrm{F}}$, which coincides with the Euclidean norm of the vector of singular values of $\bSigma$. In view of the equivalence of vector norms on 
$\Real^{p(p+1)/2}$,
\begin{eqnarray*}
\frac{\alpha}{2} \|\bSigma\|_{*} \geq \eta \|\bSigma\|_{\mathrm{F}}
\end{eqnarray*}
for some positive constant $\eta$. Integrating the resulting inequality
\begin{eqnarray*}
e^{- {\lambda \over 2}\left[\alpha \|\bSigma\|_{*}+(1-\alpha)\|\bSigma^{-1}\|_{*}\right]}
& \le & e^{-\eta \lambda \|\bSigma\|_{\mathrm{F}}}
\end{eqnarray*}
over $\bSigma$ demonstrates that the prior is proper. The normalizing constant of $\pi(\bSigma)$ is 
irrelevant in the ensuing discussion. Consider therefore minimization of the objective function
\begin{eqnarray*}
f(\bSigma) & = &  {n \over 2}\ln \det \bSigma+{n \over 2}\tr(\bS \bSigma^{-1}) + {\lambda \over 2}\left[\alpha \|\bSigma\|_{*}+(1-\alpha)\|\bSigma^{-1}\|_{*}\right] .
\end{eqnarray*}
The maximum of $-f(\bSigma)$ occurs at the posterior mode. In the limit as $\lambda$ 
tends to 0, $-f(\bSigma)$ reduces to the loglikelihood $-\ell(\bSigma)$. In the sequel we will refer
to our MAP covariance estimate by the acronym CERNN (Covariance Estimate Regularized by Nuclear Norms).

Fortunately, three of the four terms of $f(\bSigma)$ can be expressed as functions of the 
eigenvalues $e_i$ of $\bSigma$.  The trace contribution presents a greater
challenge.  As before, let $\bS = \bU \bD \bU^t$ denote the spectral decomposition of $\bS$
with nonnegative diagonal entries $d_i$ ordered from largest to smallest.  Likewise,
let $\bSigma = \bV \bE \bV^t$ denote the spectral decomposition of $\bSigma$
with positive diagonal entries $e_i$ ordered from largest to smallest.  In view of the
von Neumann-Fan inequality \citep{Mirsky1975},  we can assert that
\begin{eqnarray*}
- \tr(\bS \bSigma^{-1}) & \le & - \sum_{i=1}^p {d_i \over e_i} ,
\end{eqnarray*}
with equality if and only if $\bV = \bU$.  Consequently, we make the latter assumption
and replace $f(\bSigma)$ by
\begin{eqnarray*}
g(\bE) & = & {n \over 2}\sum_{i=1}^p \ln e_i +{n \over 2}\sum_{i=1}^p {d_i \over e_i}  
+{\lambda \over 2}\left[ \alpha \sum_{i=1}^p e_i+ (1-\alpha)\sum_{i=1}^p {1 \over e_i}\right] 
\end{eqnarray*}
using the cyclic permutation property of the trace function.  At a stationary point of 
$g(\bE)$, we have 
\begin{eqnarray*}
0 & = & {n \over e_i}-{nd_i+\lambda (1-\alpha)\over e_i^2}+ \lambda \alpha .
\end{eqnarray*}
The solution to this essentially quadratic equation is
\begin{eqnarray}
e_i & = & {-n + \sqrt{n^2+4\lambda \alpha [ n d_i+\lambda (1-\alpha)]} \over 2 \lambda \alpha} . 
\label{penalized_covariance_estimate}
\end{eqnarray}
We reject the negative root as inconsistent with $\bSigma$ being positive definite.
For the special case $n=0$ of no data, all $e_i= \sqrt{(1-\alpha)/\alpha}$, and the prior mode occurs 
at a multiple of the identity matrix. 
%
%

In contrast, CNR shrinks the sample eigenvalues $d_i$ as follows
\begin{eqnarray}
\label{eq:CNR_shrink}
e_i & = & \begin{cases}
\kappa_{\max} \tau^*& \text{$d_i \geq \kappa_{\max}\tau^*$} \\
d_i & \text{ $\tau^* < d_i < \kappa_{\max}\tau^*$} \\
\tau^* & \text{$d_i \leq \tau^*$},
\end{cases}
\end{eqnarray}
for a $\tau^* > 0$ that is determined from the data. Thus, CNR truncates extreme sample eigenvalues and leaves the moderate ones unchanged.

\begin{figure}
\centering
\includegraphics[scale=0.7125]{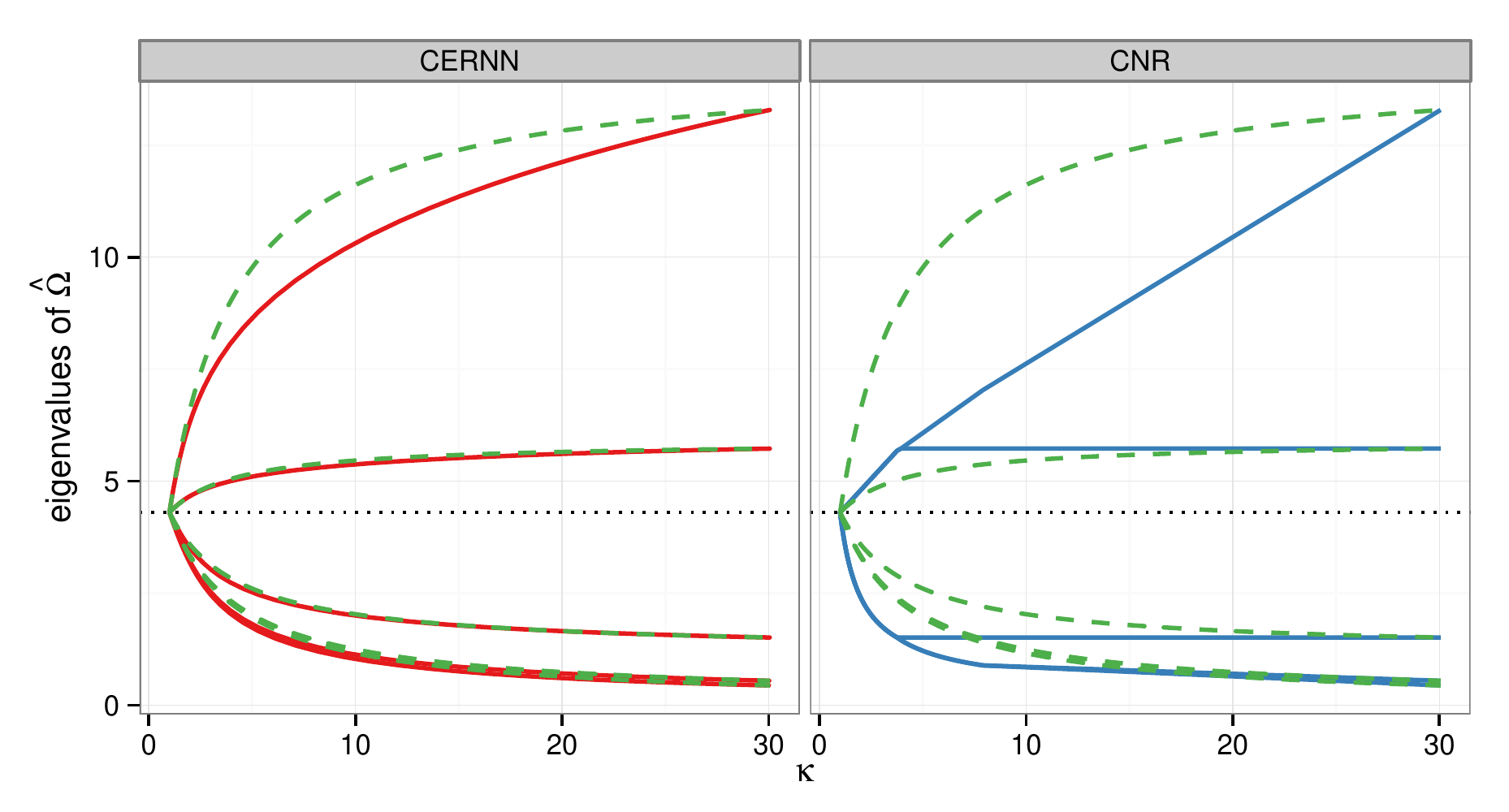}
\caption{A comparison of the solution paths for CERNN (left panel, solid line) and CNR (right panel, solid line) against the path for the linear estimator (both panels, dashed line)
for sample eigenvalues (13.29, 5.73, 1.51, 0.55, 0.44). The dotted line indicates the mean of the sample eigenvalues.
}
\label{fig:regularization_paths}
\end{figure}

Figure~\ref{fig:regularization_paths} compares the eigenvalue solution paths obtained by CERNN and CNR to 
the solution paths of the linear estimator on a set of five sample eigenvalues (13.29, 5.73, 1.51, 0.55, 0.44).
At each condition number $\kappa$, the regularization parameters ($\lambda, \kappa_{\max}, \gamma$) of the three methods are adjusted to give a condition number matching $\kappa$. Each path starts as a sample eigenvalue and ends at the mean $\hat{\sigma}$ of the sample eigenvalues. As desired, all three methods pull the large eigenvalues down towards $\hat{\sigma}$ and the small eigenvalues up towards $\hat{\sigma}$. There are important differences, however. Compared to the linear estimator, both of the non-linear estimators pull the larger eigenvalues down more aggressively and pull the smaller eigenvalues up less aggressively. The discrepancy in the treatment of high and low eigenvalues is more pronounced in CNR than in CERNN. We will see later in finite sample experiments that this more moderate approach usually leads to better performance.

Some insight to the limiting behavior of our estimator can be gained through asymptotic expansions.
Holding all but one variable fixed in formula (\ref{penalized_covariance_estimate}), 
one can demonstrate after a fair amount of algebra that  
\begin{eqnarray}
e_i  & = &  d_i+{\lambda(1-\alpha - \alpha d_i^2) \over n}+O\left({1 \over n^2}\right), \quad n \to \infty 
\label{eq:asymexp} \\
e_i  & =  & \sqrt{{1-\alpha \over \alpha}}+
\left[\sqrt{{1- \alpha \over \alpha}}{nd_i \over 2(1-\alpha)} -{n \over 2 \alpha}\right]{1 \over \lambda} 
+ O\left({1 \over \lambda^2}\right), \quad \lambda \to \infty. \nonumber
\end{eqnarray}
These asymptotic expansions accord with common sense.  Namely, the data eventually overwhelms
a fixed prior, and increasing the penalty strength for a fixed amount of data pulls the estimate of $\bSigma$
toward the prior mode. The second asymptotic expansion will prove useful in selecting $\lambda$ in a practical data-dependent way.

\subsection{Choosing $\lambda$ and $\alpha$}
\label{sec:crossvalidation}
Let us first consider how to choose $\alpha$. A natural choice for it would match the prior to the scale of the data. Thus, we would determine $\alpha$ as the 
solution to the equation
\begin{eqnarray*}
p \sqrt{{1-\alpha \over \alpha}} & = & \tr\left( \sqrt{{1-\alpha \over \alpha}} \bI \right)
\:\;\, = \:\;\, \tr(\bS),
\end{eqnarray*}
namely
\begin{eqnarray}
\hat{\alpha} = \left [ 1 + \left ( \frac{1}{p} \tr (\bS) \right )^2 \right ]^{-1}. \label{alpha_estimate}
\end{eqnarray}
Of course, we recognize that this choice of $\hat{\alpha}$ results in the prior mode being $\hat{\sigma}\bI$. For the remainder of the paper we will set $\alpha$ to $\hat{\alpha}$.

We recommend different strategies for choosing $\lambda$ depending on whether the covariance estimation is being performed for a supervised or unsupervised task. Both strategies employ cross-validation. We defer describing the former strategy until we discuss an application to discriminant analysis. To choose $\lambda$ in the unsupervised context, we implement cross-validation as follows. Let $\bY \in \Real^{n \times p}$ denote the observed data. In $K$-fold cross validation we partition the observations, or rows of $\bY$, into $K$ disjoint sets. Let $\bY_k$ denote the $k$th subset, $n_k$ denote the number of its rows, and $\hat{\bSigma}_\lambda^{(-k)}$ denote the estimate we obtain when we fit the data using all but the $k$th partition $\bY_k$. We have indicated with our notation that our estimate depends on $\lambda$ but not $\alpha$ since we have fixed $\hat{\alpha}$ at the value displayed in equation (\ref{alpha_estimate}). For a grid of increasing regularization parameters, $\lambda = 0, \ldots, \lambda_{\max}$, we evaluate the predictive negative loglikelihood of $\hat{\bSigma}_\lambda^{(-k)}$ on the held out $k$th subgroup $\bY_k$
\begin{eqnarray*}
\ell_k(\hat{\bSigma}_\lambda^{(-k)},\bY_k) = \frac{n_k}{2}  \ln \det \hat{\bSigma}_\lambda^{(-k)} + \frac{n_k}{2} \tr \left ( \frac{1}{n_k} \bY_k^t\bY_k \left [ \hat{\bSigma}_\lambda^{(-k)} \right]^{-1}\right ).
\end{eqnarray*}
We select the $\lambda$ that minimizes the average $\ell_k$ over the $K$ folds, namely
\begin{eqnarray*}
\hat{\lambda} = \underset{\lambda \in \{0, \ldots, \lambda_{\max}\}}{\arg\min}\; \frac{1}{n} \sum_{k=1}^K \ell_k(\hat{\bSigma}_\lambda^{(-k)}, \bY_k).
\end{eqnarray*}
We want to choose $\lambda_{\max}$ sufficiently large to adequately explore the dynamic range of possible solutions. The second expansion in
(\ref{eq:asymexp}) gives us a rough bound on how much each of the shrunken eigenvalues deviates from the prior mode. We choose $\lambda_{\max}$ so that those
deviations are no more than some small fraction $\varepsilon$ of the prior mode, namely we choose $\lambda_{\max}$ so that
\begin{eqnarray*}
\underset{i = 1, \ldots, p}{\max}\; \left \lvert \sqrt{{1- \hat{\alpha} \over \hat{\alpha}}}{nd_i \over 2(1-\hat{\alpha})} -{n \over 2 \hat{\alpha}}\right \rvert {1 \over {\lambda_{\max}}} \leq \varepsilon \sqrt{\frac{1-\hat{\alpha}}{\hat{\alpha}}} ,
\end{eqnarray*}
where $\varepsilon \ll 1$, say $10^{-2}$.

\section{Consistency and Asymptotic Efficiency}
\label{sec:asymptotics}

In proving consistency, we will need various facts.  First, suppose $\bA$ and $\bB$ 
are two $p \times p$ symmetric matrices with ordered eigenvalues $\{a_i\}_{i=1}^p$ 
and $\{b_i\}_{i=1}^p$.  Then one has
\begin{eqnarray}
\sum_{i=1}^p (a_i-b_i)^2 & \le & \|\bA-\bB\|_{\mathrm{F}}^2. \label{Frobenius-bound}
\end{eqnarray}
This is a consequence of Fan's inequality because $\sum_{i=1}^p a_i^2 = \|\bA\|_{\mathrm{F}}^2$ 
and $\sum_{i=1}^p b_i^2 = \|\bB\|_{\mathrm{F}}^2$.  If the two matrices $\bA = \bU \diag(\ba) \bU^t$ 
and $\bB = \bU \diag(\bb) \bU^t $ are simultaneously diagonalizable, then equality holds in inequality
(\ref{Frobenius-bound}).  We will also need the inequalities
\begin{eqnarray}
\sqrt{1+x} & \le & 1+{x \over 2} \quad \mbox{and} \quad \sqrt{1+x} \:\;\, \ge \:\;\,
 1+{x \over 2} -{x^2 \over 8} \label{completely_montonic}
 \end{eqnarray}
for nonnegative $x$.  Verification will be left to the reader based on the fact that the derivatives
of $\sqrt{1+x}$ alternate in sign.  Functions having this property are said to be completely monotonic.

Let $\bS_n$ be the sample covariance matrix with eigenvalues $d_{n1}$ through $d_{np}$
for the first $n$ sample points.  The sequence $\bS_n$ converges almost surely to the true covariance 
matrix $\bSigma$ with eigenvalues $\omega_1$ through $\omega_p$.  Inequality (\ref{Frobenius-bound})
therefore implies $\lim_{n \to \infty} \sum_{i=1}^p (d_{ni}-\omega_i)^2 = 0$.  On this basis we
will argue that $\lim_{n \to \infty} \sum_{i=1}^p (e_{ni}-\omega_i)^2 = 0$ as well, where the
$e_{ni}$ are the transformed eigenvalues of $\bS_n$.  To make this reasoning rigorous,
we must show that the asymptotic expansion (\ref{eq:asymexp})
is uniform as the eigenvalues $d_{ni}$ converge to the eigenvalues $\omega_i$.  This is 
where the inequalities (\ref{completely_montonic}) come into play.  Indeed, we have
\begin{eqnarray}
{\lambda (1-\alpha) \over n}-{n \over 2 \lambda \alpha}{x^2 \over 8} 
& \le & e_{ni} -d_{ni} \:\;\, \le \:\;\, {\lambda (1-\alpha) \over n}
\label{eigen-bouund} \\
x & = & {4 \lambda \alpha d_{ni} \over n}+{4 \lambda^2 \alpha(1-\alpha) \over n^2} .
\nonumber
\end{eqnarray}
The identity
\begin{eqnarray*}
\|\bS_n - \bSigma_n\|_{\mathrm{F}}^2 & = & \sum_{i=1}^p (d_{ni}-e_{ni})^2 
\end{eqnarray*}
finishes the proof that $\bSigma_n$ tends to $\bSigma$. 

Now consider the question of asymptotic efficiency.  The scaled difference
$\sqrt{n}(\bS_n-\bSigma)$ tends in distribution to a multivariate normal
distribution with mean  $\bf 0$ because the sequence of estimators
$\bS_n$ is asymptotically efficient \citep{Fer1996}. The representation
\begin{eqnarray*}
\sqrt{n}(\bSigma_n-\bSigma) & = & \sqrt{n}(\bS_n-\bSigma)+\sqrt{n}(\bSigma_n-\bS_n)
\end{eqnarray*}
and Slutsky's theorem \citep{Fer1996} imply that $\sqrt{n}(\bSigma_n-\bSigma)$ 
tends in distribution to the same limit. In this regard note that
\begin{eqnarray*}
\|\sqrt{n}(\bSigma_n-\bS_n)\|_{\mathrm{F}}^2 & = & n \sum_{i=1}^p (d_{ni}-e_{ni})^2 
\end{eqnarray*}
tends almost surely to 0 owing to the bounds (\ref{eigen-bouund}) and 
the convergence of $d_{ni}$ to $\omega_i$.

CNR and linear estimators are also asymptotically very well behaved. The asymptotic properties of CNR are stated in terms of 
the entropy risk, which is the expectation of the entropy loss given below
\begin{eqnarray}
\label{eq:loss_entropy}
\mathcal{L}_{\text{E}}(\bSigma_n,\bSigma) = \tr(\bSigma^{-1}\bSigma_n) - \log \det (\bSigma^{-1}\bSigma_n) - p.
\end{eqnarray}
CNR has asymptotically lower entropy risk than the sample covariance \citep{WonLimKim2012}. But since the sample covariance is a consistent estimator of the covariance, it follows that CNR is also a consistent estimator.

Among all linear estimators, the LW estimator of \cite{LedWol2004} is asymptotically optimal with respect to a quadratic risk. To make this claim more precise, consider the optimization problem,
\begin{eqnarray}
\label{eq:lw_opt}
&\underset{\gamma_1, \gamma_2}{\min}& \; \| \tilde{\bSigma} - \bSigma \|_{\mathrm{F}}^2 \\
&\text{subject to}& \tilde{\bSigma} = \gamma_1\bI + \gamma_2 \bS_n, \nonumber
\end{eqnarray}
where the weights $\gamma_1$ and $\gamma_2$ are allowed to be random and can therefore depend on the data. One can think of the solution $\bSigma^\star$ to (\ref{eq:lw_opt}) as being the best possible linear combination of $\bI$ and the sample covariance $\bS_n$. Even though $\bSigma^\star$ may not be an estimator, since it is allowed to depend on the unseen population covariance $\bSigma$, \cite{LedWol2004} prove that the quadratic risk of LW has the same quadratic risk as $\bSigma^\star$ asymptotically. 
Their results are actually even stronger than stated here because $p$ is allowed to increase along with $n$, under suitable, but somewhat complicated regularity conditions that we omit. Since the sample covariance is consistent, and its quadratic risk by definition exceeds the quadratic risk of $\bSigma^\star$, it follows that LW is also a consistent estimator.

\section{Finite Sample Performance}
\label{sec:comparisons}

Given their similar asymptotic behavior, to better understand the relative strengths of CERNN, CNR, and the optimal linear estimator, LW, we conducted a finite sample study almost identical to the one carried out by \cite{WonLimKim2012}.
We assessed the estimators based on two commonly used criteria, namely the entropy loss (\ref{eq:loss_entropy})
and the quadratic loss
\begin{eqnarray*}
\mathcal{L}_{\text{Q}}(\hat{\bSigma},\bSigma) = \lVert \hat{\bSigma}\bSigma^{-1} - \bI \rVert^2_{\mathrm{F}}.
\end{eqnarray*}
In our experiments, we simulated data in which we varied the ratio $p/n$ and the spread of the eigenvalues of the population covariance. As noted earlier, linear shrinkage improves on the sample covariance when $p/n$ is large or the population eigenvalues are close to one another. Won et al.\@ report that when the eigenvalues of the population covariance are bimodal, CNR dramatically outperforms linear estimators if very few eigenvalues take on the high values. As the proportion of large eigenvalues grows, however, the discrepancy diminishes. CERNN shrinks extreme eigenvalues in a manner similar to CRN but less drastically and shrinks intermediate eigenvalues similarly to linear estimators. In contrast CNR leaves intermediate eigenvalues unchanged. Consequently, we anticipate that CERNN has the potential to improve on CNR in the latter case, where there is a need to shrink all eigenvalues, like linear estimators, but with extra emphasis on the extreme ones, unlike linear estimators.

We simulated $p$-dimensional zero mean multivariate normal samples with diagonal covariances with $p = 120$, 250, and 500. The eigenvalues took on either high values of $1 - \upsilon + \upsilon p$ or low values 
of $1 - \upsilon$ where $\upsilon = 0.1$. For each $p$, the number of high value eigenvalues ranged from a single high value to 10\%, 20\%, 30\%, and 40\% of $p$. For each $p$ we drew one of three sample sizes $n$ such that the ratio $r=p/n$ took on values that were roughly 1.25, 2, or 4. For each choice of $p$, $n$, and $r$, we simulated 100 data sets and computed CNR, CERNN, and LW. For each data set, we chose an optimal $\kappa_{\max}$ and $\lambda$ via 10-fold cross-validation. We used the R package \textbf{CondReg} to obtain the CNR estimate \citep{OhRaj2013}. LW specifies a penalization parameter $\gamma$ based on the data.

To expedite comparisons, we report the average ratio of the loss of other estimators to the loss of CERNN. When this ratio is less than one, the other estimator is performing better than CERNN, and when the ratio is greater than one, CERNN is performing better. Tables~\ref{tab:loss_quadratic} and \ref{tab:loss_entropy} summarize comparisons under the quadratic loss and entropy loss respectively.

We first note that these experiments confirm what was previously observed by \cite{WonLimKim2012}. Regardless of loss criterion, CNR typically outperforms linear estimators over a wide range of scenarios, but especially when few eigenvalues dominate and the ratio $p/n$ is larger. Compared to CERNN, however, CNR soundly outperforms CERNN only in the extreme case of a single large eigenvalue. In all other scenarios, under either loss criterion, CERNN performs better. It is especially notable that CERNN performs very well in comparison to CNR in scenarios where CNR provides marginal improvement over linear estimators, namely when the fraction of high eigenvalues is highest at 40\%. 
According to equation (\ref{eq:CNR_shrink}), recall that CNR only shrinks the most extreme sample eigenvalues and leaves the intermediate  eigenvalues unchanged. It is not surprising then that it works best when there are very few large population eigenvalues and loses its competitive edge in the opposite circumstance. CERNN's more moderate approach of shrinking all eigenvalues, with extra emphasis on larger ones, appears to win out when there is a more balanced mix of high and low eigenvalues.

\begin{table}
\centering
\begin{tabular}{rrrrrrrrrr}
\toprule
\multicolumn{7}{c}{singleton} \\
\cmidrule(lr){1-7} 
& \multicolumn{3}{c}{CNR/CERNN} & \multicolumn{3}{c}{LW/CERNN}   \\
$p = $ & $125$  & $250$  & $500$ & $125$  & $250$  & $500$ \\
\cmidrule(lr){1-7} 
$p/n=4$ & {\bf 0.42} (0.13) & {\bf 0.19} (0.03) & {\bf 0.12} (0.01) & 12.4 (6.05) & 17.9 (3.97) & 33.2 (4.23) \\ 
$2$ & {\bf 0.36} (0.07) & {\bf 0.23} (0.03) & {\bf 0.20} (0.02) & 9.28 (2.76) & 18.1 (2.66) & 31.7 (2.09) \\ 
$1.25$ & {\bf 0.37} (0.04) & {\bf 0.29} (0.03) & {\bf 0.27} (0.02) & 8.87 (1.59) & 16.9 (1.48) & 26.7 (1.18) \\ 
\end{tabular}
\begin{tabular}{rrrrrrrrrr}
\toprule
\multicolumn{7}{c}{$10\%$ high} \\
\cmidrule(lr){1-7} 
& \multicolumn{3}{c}{CNR/CERNN} & \multicolumn{3}{c}{LW/CERNN}  \\
$p = $ & $125$  & $250$  & $500$ & $125$  & $250$  & $500$  \\
\cmidrule(lr){1-7} 
$p/n=4$ & 1.15 (0.20) & 1.80 (0.12) & 2.77 (0.13) & 4.91 (0.53) & 8.01 (0.42) & 17.0 (0.66) \\ 
$2$ & 1.18 (0.11) & 1.66 (0.08) & 1.84 (0.06) & 4.42 (0.30) & 6.85 (0.28) & 12.3 (0.41) \\ 
$1.25$ & 1.46 (0.08) & 1.39 (0.05) & 1.81 (0.05) & 4.14 (0.21) & 6.08 (0.20) & 10.3 (0.29) \\ 
\end{tabular}
\begin{tabular}{rrrrrrrrrr}
\toprule
\multicolumn{7}{c}{$20\%$ high} \\
\cmidrule(lr){1-7} 
& \multicolumn{3}{c}{CNR/CERNN} & \multicolumn{3}{c}{LW/CERNN}  \\
$p = $ & $125$  & $250$  & $500$ & $125$  & $250$  & $500$ \\
\cmidrule(lr){1-7} 
$p/n=4$ & 4.28 (0.70) & 8.52 (0.86) & 18.5 (1.06) & 8.72 (0.88) & 23.5 (1.62) & 78.5 (3.32) \\ 
2 & 1.97 (0.16) & 2.51 (0.13) & 5.25 (0.16) & 6.10 (0.37) & 12.9 (0.53) & 31.1 (1.00) \\ 
1.25 & 1.45 (0.07) & 1.95 (0.07) & 4.90 (0.15) & 4.98 (0.20) & 9.42 (0.31) & 20.9 (0.61) \\ 
\end{tabular}
\begin{tabular}{rrrrrrrrrr}
\toprule
\multicolumn{7}{c}{$30\%$ high} \\
\cmidrule(lr){1-7} 
& \multicolumn{3}{c}{CNR/CERNN} & \multicolumn{3}{c}{LW/CERNN}  \\
$p = $ & $125$  & $250$  & $500$ & $125$  & $250$  & $500$ \\
\cmidrule(lr){1-7} 
$p/n=4$ & 11.9 (1.90) & 37.8 (3.17) & 128 (6.03) & 16.9 (1.74) & 62.6 (3.49) & 244 (7.32) \\ 
2 & 3.97 (0.36) & 6.65 (0.34) & 13.6 (0.31) & 10.7 (0.67) & 31.4 (1.24) & 95.0 (2.22) \\ 
1.25 & 2.15 (0.10) & 3.76 (0.12) & 10.6 (0.30) & 7.71 (0.36) & 18.6 (0.63) & 51.8 (1.56) \\ 
\end{tabular}
\begin{tabular}{rrrrrrrrrr}
\toprule
\multicolumn{7}{c}{$40\%$ high} \\
\cmidrule(lr){1-7} 
& \multicolumn{3}{c}{CNR/CERNN} & \multicolumn{3}{c}{LW/CERNN}  \\
$p = $ & $125$  & $250$  & $500$ & $125$  & $250$  & $500$ \\
\cmidrule(lr){1-7} 
$p/n=4$ & 25.6 (3.18) & 83.2 (5.06) & 304 (10.7) & 28.3 (2.54) & 105 (4.60) & 407 (10.9) \\ 
2 & 11.0 (1.10) & 28.2 (1.61) & 68.9 (3.48) & 20.1 (1.23) & 66.9 (2.28) & 234 (4.74) \\ 
1.25 & 3.66 (0.22) & 6.39 (0.14) & 18.9 (0.27) & 13.5 (0.68) & 38.6 (1.05) & 119 (1.62) \\ 
\bottomrule
\end{tabular}
\caption{Average ratio of quadratic loss of CNR and LW to that of CERNN. Values less than 1 indicate better performance than CERNN (bold). Standard deviations are in parentheses.}
\label{tab:loss_quadratic}
\end{table}

\begin{table}
\centering
\begin{tabular}{rrrrrrrrrr}
\toprule
\multicolumn{7}{c}{singleton} \\
\cmidrule(lr){1-7} 
& \multicolumn{3}{c}{CNR/CERNN} & \multicolumn{3}{c}{LW/CERNN}   \\
$p = $ & $125$  & $250$  & $500$ & $125$  & $250$  & $500$ \\
\cmidrule(lr){1-7} 
$p/n=4$ & {\bf 0.83} (0.11) & {\bf 0.56} (0.05) & {\bf 0.40} (0.03) & 5.38 (2.24) & 9.20 (2.10) & 18.7 (2.61) \\ 
  2 & {\bf 0.78} (0.07) & {\bf 0.59} (0.03) & {\bf 0.49} (0.02) & 5.24 (1.63) & 11.9 (2.08) & 24.6 (2.17) \\ 
  1.25 & {\bf 0.79} (0.04) & {\bf 0.66} (0.03) & {\bf 0.56} (0.02) & 5.75 (1.15) & 12.8 (1.37) & 24.4 (1.30) \\ 
\end{tabular}
\begin{tabular}{rrrrrrrrrr}
\toprule
\multicolumn{7}{c}{$10\%$ high} \\
\cmidrule(lr){1-7} 
& \multicolumn{3}{c}{CNR/CERNN} & \multicolumn{3}{c}{LW/CERNN}  \\
$p = $ & $125$  & $250$  & $500$ & $125$  & $250$  & $500$  \\
\cmidrule(lr){1-7} 
$p/n=4$& 1.11 (0.12) & 1.34 (0.07) & 1.62 (0.06) & 1.43 (0.11) & 1.83 (0.09) & 3.19 (0.11) \\ 
$2$ & 1.04 (0.07) & 1.08 (0.04) & 1.16 (0.03) & 1.29 (0.07) & 1.34 (0.05) & 1.81 (0.05) \\ 
$1.25$ & 1.13 (0.05) & 0.96 (0.03) & 1.26 (0.03) & 1.29 (0.05) & 1.20 (0.03) & 1.38 (0.03) \\ 
\end{tabular}
\begin{tabular}{rrrrrrrrrr}
\toprule
\multicolumn{7}{c}{$20\%$ high} \\
\cmidrule(lr){1-7} 
& \multicolumn{3}{c}{CNR/CERNN} & \multicolumn{3}{c}{LW/CERNN}  \\
$p = $ & $125$  & $250$  & $500$ & $125$  & $250$  & $500$ \\
\cmidrule(lr){1-7} 
$p/n=4$ & 1.90 (0.19) & 2.71 (0.15) & 3.74 (0.10) & 2.16 (0.19) & 4.16 (0.20) & 8.27 (0.17) \\ 
$2$ & 1.19 (0.06) & 1.20 (0.05) & 1.97 (0.03) & 1.73 (0.10) & 2.93 (0.10) & 5.57 (0.11) \\ 
$1.25$ & {\bf 0.92} (0.03) & 1.03 (0.02) & 2.14 (0.04) & 1.42 (0.05) & 2.18 (0.06) & 4.05 (0.07) \\ 
\end{tabular}
\begin{tabular}{rrrrrrrrrr}
\toprule
\multicolumn{7}{c}{$30\%$ high} \\
\cmidrule(lr){1-7} 
& \multicolumn{3}{c}{CNR/CERNN} & \multicolumn{3}{c}{LW/CERNN}  \\
$p = $ & $125$  & $250$  & $500$ & $125$  & $250$  & $500$ \\
\cmidrule(lr){1-7} 
$p/n=4$ & 2.50 (0.20) & 4.04 (0.17) & 6.49 (0.15) & 2.69 (0.17) & 5.06 (0.17) & 9.06 (0.15) \\ 
$2$ & 1.50 (0.08) & 1.72 (0.06) & 2.39 (0.02) & 2.44 (0.12) & 4.58 (0.12) & 8.53 (0.10) \\ 
$1.25$ & {\bf 0.96} (0.03) & 1.24 (0.02) & 2.64 (0.02) & 2.11 (0.10) & 3.78 (0.10) & 7.13 (0.07) \\ 
\end{tabular}
\begin{tabular}{rrrrrrrrrr}
\toprule
\multicolumn{7}{c}{$40\%$ high} \\
\cmidrule(lr){1-7} 
& \multicolumn{3}{c}{CNR/CERNN} & \multicolumn{3}{c}{LW/CERNN}  \\
$p = $ & $125$  & $250$  & $500$ & $125$  & $250$  & $500$ \\
\cmidrule(lr){1-7} 
$p/n=4$ & 2.80 (0.17) & 4.44 (0.13) & 7.29 (0.13) & 2.79 (0.15) & 4.91 (0.12) & 8.45 (0.12) \\ 
$2$ & 2.09 (0.11) & 3.04 (0.11) & 4.23 (0.12) & 2.81 (0.13) & 4.91 (0.12) & 8.51 (0.09) \\ 
$1.25$ & 1.13 (0.05) & 1.33 (0.02) & 2.66 (0.02) & 2.67 (0.11) & 4.79 (0.09) & 8.64 (0.08) \\ 
\bottomrule
\end{tabular}
\caption{Average ratio of entropy loss of CNR and LW to that of CERNN. Values less than 1 indicate better performance than CERNN (bold). Standard deviations are in parentheses.}
\label{tab:loss_entropy}
\end{table}

%

\section{Applications}
\label{sec:applications}

Several common statistical procedures are potential beneficiaries of shrinkage estimation of sample covariance matrices. Here we illustrate how CERNN applies to discriminant analysis and clustering.

\subsection{Discriminant Analysis}
\label{sec:discriminant}

The classical discriminant function 
\begin{eqnarray*}
\delta_k(\bx) & = & \bx^t \bSigma^{-1}\bmu_k- \bmu_k^t \bSigma^{-1}\bmu_k+\ln \pi_k,
\end{eqnarray*}
incorporates the mean $\bmu_k$ and prior probability $\pi_k$ of each
class $k$. A new observation $\bx$ is assigned to the class $k$ maximizing
$\delta_k(\bx)$. If there are $c$ classes ${\cal C}_1,\ldots,{\cal C}_c$, 
then the standard estimator of $\bSigma$ is
\begin{eqnarray*}
\hat{\bSigma} & = & \frac{1}{n-c}\sum_{k=1}^c \sum_{i \in {\cal C}_k} 
(\bx_i-\hat{\bmu}_k)(\bx_i-\hat{\bmu}_k)^t ,
\end{eqnarray*}
where
\begin{eqnarray*}
\hat{\bmu}_k & = & \frac{1}{|{\cal C}_k|}\sum_{i \in {\cal C}_k} \bx_i .
\end{eqnarray*}
One can obviously shrink $\hat{\bSigma}$ 
to moderate its eigenvalues.   In quadratic discriminant analysis,
a separate covariance matrix $\bSigma_k$ is assigned to each class $k$. These 
are estimated in the usual way, and eigenvalue shrinkage is likely even more
beneficial than in linear discriminant analysis. \cite{Fri1989} advocates
regularized discriminant analysis (RDA), a compromise between 
linear and quadratic discriminant analysis that shrinks 
$\bSigma_k$ toward a common $\bSigma$ via a convex combination $\gamma \bSigma_k + (1-\gamma) \bSigma$. 
Although Friedman also suggests shrinking toward class specific multiples of the identity matrix, we do not consider his more complicated version here. \cite{Guo2007} shrink covariance estimates towards the identity matrix and also
apply lasso shrinkage on the centroids to obtain improved classification performance in microarray studies. The main difference between CERNN and these methods is that CERNN performs nonlinear shrinkage of the sample eigenvalues.

Since we are primarily interested in the case where all or most of the predictors are instrumental in grouping, we consider only
Friedman's method in a comparison on three data sets from the UCI machine learning repository \citep{BacLic2013}. In the case of the E. Coli data set, we restricted analysis to the five most abundant classes. We split each data set into training and testing sets. In each experiment we used 1/5 of the data for training and 4/5 for testing. Table~\ref{tab:rda} records the number of samples per group in each set. In these data poor examples, even linear discriminant analysis is not viable since a common sample covariance estimate will be ill-conditioned if not singular. Nonetheless, our results show that the combination of separate covariances with regularization works well. We modeled a separate covariance for each class and used 5-fold cross validation to select $c$ regularization parameters for CERNN. We used essentially the same procedure described in \ref{sec:crossvalidation} except that we used the misclassification rate instead of the predictive negative loglikelihood.
We also used 5-fold cross validation to select the single $\gamma$ parameter for \citep{Fri1989}. The testing errors in Table~\ref{tab:rda} demonstrate that CERNN performs well in comparison with RDA.  Even when it does not perform as accurately, its drop off is small.

\begin{table}[htb] \centering 
  \caption{Comparison of CERNN and RDA on three data sets from the UCI machine learning repository. The fourth column
indicates the number of parameters (mean and covariance) per group in the QDA model. The fifth and sixth columns breakdown the number of samples per group. The last two columns report the classification success rate in the test set.} 
  \label{tab:rda} 
\footnotesize 

\begin{tabular}{@{\extracolsep{5pt}} c c c c c c c c } 
\\[-1.8ex]\hline 
\hline \\[-1.8ex] 
data & $p$ & $c$ & $\frac{p(p+3)}{2}$  & samples (train) & samples (test) & CERNN & RDA \\
\hline \\[-1.8ex] 
wine &  $13$ & $3$ & $104$ & 13/13/10 & 46/58/38 & $0.859$ & $0.627$ \\ 
seeds & $7$ & $3$ & $35$ & 14/15/13 & 56/55/57 & $0.929$ & $0.935$ \\ 
ecoli & $7$ & $5$ & $35$ & 30/17/7/3/9 & 113/60/28/17/43 & $0.670$ & $0.705$ \\ 
\hline \\[-1.8ex] 
\normalsize 
\end{tabular} 
\end{table} 

\subsection{Covariance Regularized EM Clustering}
\label{sec:clustering}

We now show how CERNN stabilizes estimation in the standard EM clustering algorithm \citep{McL2000}.
Let $\phi(\by \mid \bmu, \bSigma)$ denote a multivariate Gaussian density with mean $\bmu$ and covariance $\bSigma$.
EM clustering revolves around the mixture density 
\begin{eqnarray*}
h(\by \mid \Xi) & = & \sum_{k=1}^c \pi_k \,\phi(\by \mid \bmu_k, \bSigma_k) 
\end{eqnarray*}
with parameters $\Xi = \{\bmu_k, \bSigma_k, \pi_k\}_{k=1}^c$. The $\pi_k$ are nonnegative mixture weights summing to 1.
We are given $n$ independent observations $\by_1, \ldots, \by_n$ from the mixture density and wish to estimate $\Xi$ by maximizing the loglikelihood. It is well known that the loglikelihood is unbounded from above and plagued by many suboptimal local extrema \citep{McL2000}. \cite{Hat1985} proposed constraints leading to a maximum likelihood problem free of singularities and beset by fewer local extrema. Later it was shown that these constraints could be met by imposing bounds on the largest and smallest eigenvalues of the $\bSigma_k$ \citep{Ing2004,IngRoc2007}. These findings suggest that employing our Bayesian prior to estimate $\bSigma_k$ can also improve the search for good local optima, since it shrinks the extreme sample eigenvalues to the sample mean eigenvalue.

If $z_{ik}$ is the indicator function of the event that observation $i$ comes from cluster $k$, then 
the complete data loglikelihood plus logprior amounts to
\begin{eqnarray*}
\ell(\Xi) = \sum_{i=1}^n \sum_{k=1}^c z_{ik} \left [ \ln \pi_k + \ln \phi(\by_i \mid \bmu_k, \bSigma_k) \right ] 
- \frac{\lambda}{2} \left[\alpha_k \|\bSigma_k\|_{*}+(1-\alpha_k)\|\bSigma_k^{-1}\|_{*}\right].
\end{eqnarray*}
Straightforward application of Bayes rule yields the conditional expectation 
\begin{eqnarray*}
w_{ik} & = & E[z_{ik} \mid \bY, \Xi] \:\;\, = \:\;\,
\frac{\pi_k \phi(\by_i \mid \bmu_k, \bSigma_k)}{\sum_{l=1}^c \pi_l \phi(\by_i \mid \bmu_l, \bSigma_l)}.
\end{eqnarray*}
These weights should be subscripted by the current iteration number $m$, but to avoid clutter we omit the subscripts.
If we set 
\begin{eqnarray*}
w_k & = & \sum_{i=1}^n w_{ik} \quad \text{and} \quad
\bS_k \:\;\, = \:\;\, \frac{1}{w_k} \sum_{i=1}^n w_{ik} (\by_i - \bmu_k)(\by_i - \bmu_k)^t,
\end{eqnarray*}
then the EM updates are $\pi_k = \frac{w_k}{n}$, $\bmu_k = \frac{1}{w_k}\sum_{i=1}^n w_{ik}\by_i$, and
\begin{eqnarray*}
\bSigma_k \, = \,  \underset{\bSigma}{\arg\min} \; \frac{w_k}{2}\log \det \bSigma + \frac{w_k}{2}\tr (\bSigma^{-1} \bS_k) + \frac{\lambda}{2} \left[\alpha_k \|\bSigma\|_{*}+(1-\alpha_k)\|\bSigma^{-1}\|_{*}\right].
\end{eqnarray*}

We now address two practical issues. First, there is the question of how to choose $\alpha_k$. In the previous examples we sought a stable estimate of a single covariance matrix. Here we seek $c$ covariance matrices whose imputed data change from iteration to iteration. In accord with our previous choice for $\alpha$, we set $\alpha_k$ to be $(1/p)\tr(\bS_k)$. This $\alpha_k$ changes dynamically as $\bS_k$ changes. Second, it is possible for $w_{ik} \approx 0$ for all $i$ for a given $k$ at some point as the algorithm iterates. Indeed, finite machine precision may assign $w_{k}$ the value 0 at some point, making the updates for $\bSigma_k$ and $\bmu_k$ undefined. Consequently, we only update $\bSigma_k$ and $\bmu_k$ when $w_{k} > 0$. This is not a major issue, however, since this scenario only arises when no data points should be assigned to the $k$th cluster.

Besides the work of \cite{IngRoc2007}, similar approaches have been employed previously. \cite{FraRaf2007} suggest a restricted parameterization of covariance matrices. While they offer a menu of parameterizations that cover a range of degrees of freedom, each model has a fixed number of degrees of freedom. One advantage of our model is that the degrees of freedom may be tuned continuously with a single parameter $\lambda$.

\begin{figure}
\centering
\begin{tabular}{c}
\subfloat[$\lambda = 0.1$]{\label{fig:cremclust0_1}
\includegraphics[scale=0.525]{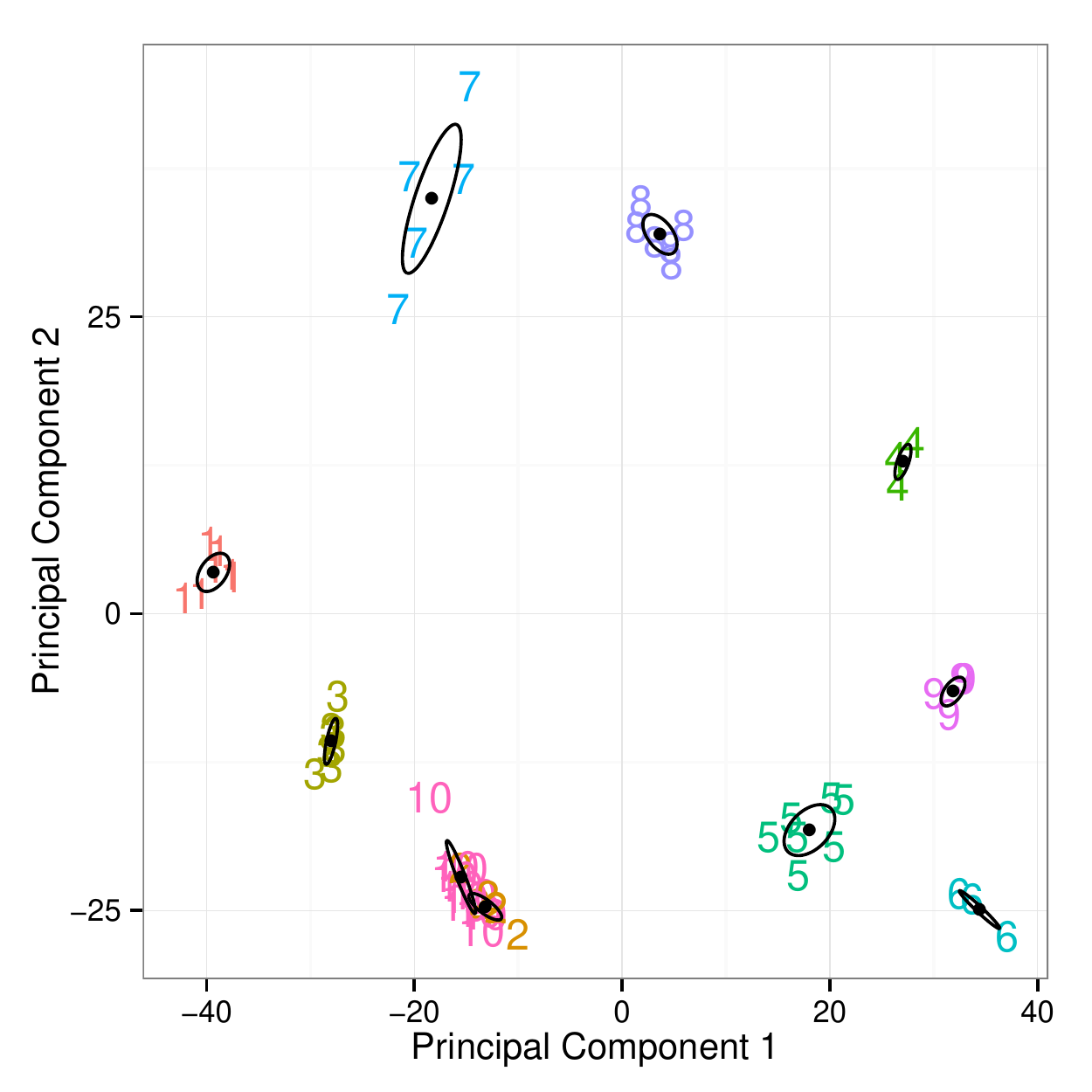}}
\subfloat[$\lambda = 10$]{\label{fig:cremclust10}
\includegraphics[scale=0.525]{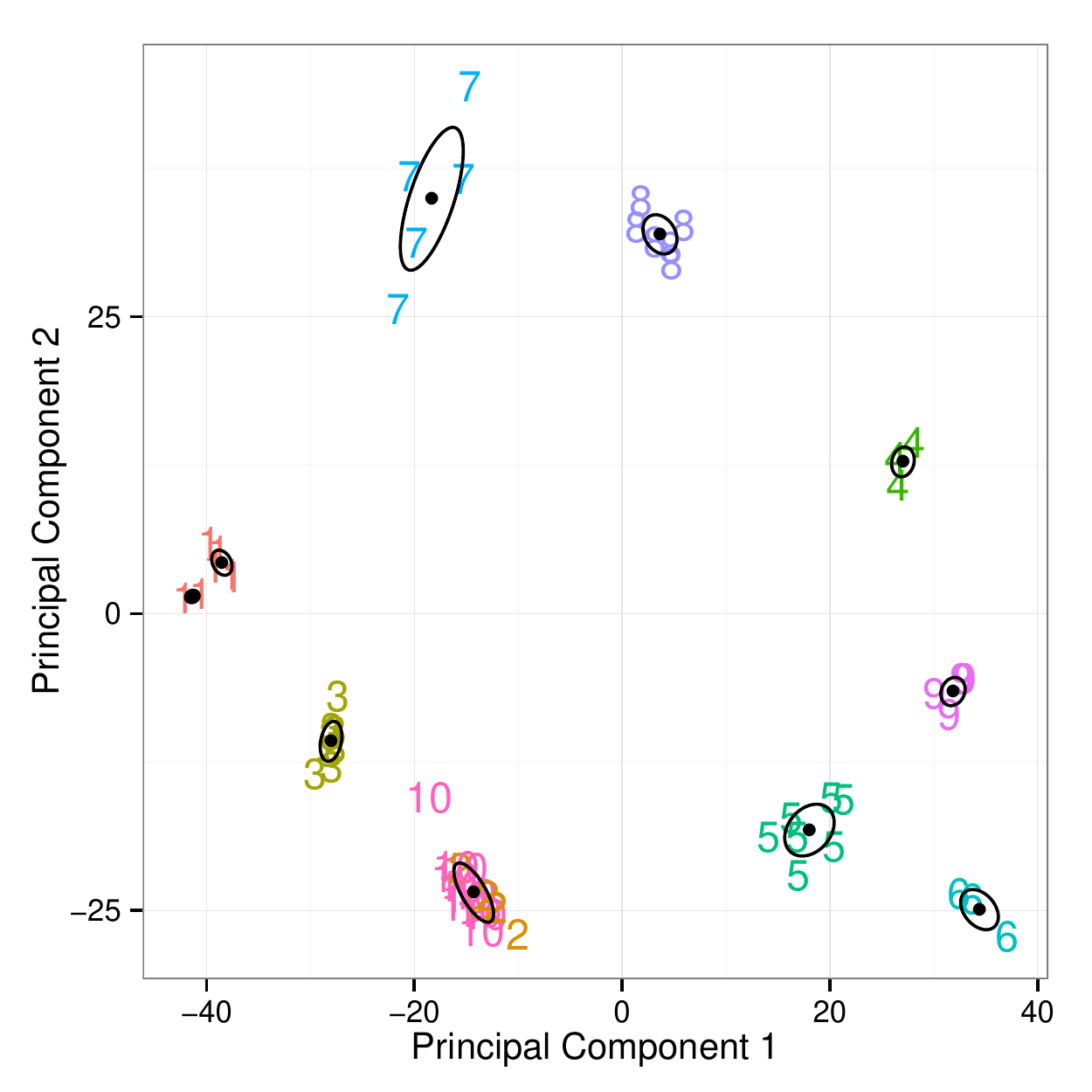}} \\
\subfloat[$\lambda = 100$]{\label{fig:cremclust100}
\includegraphics[scale=0.525]{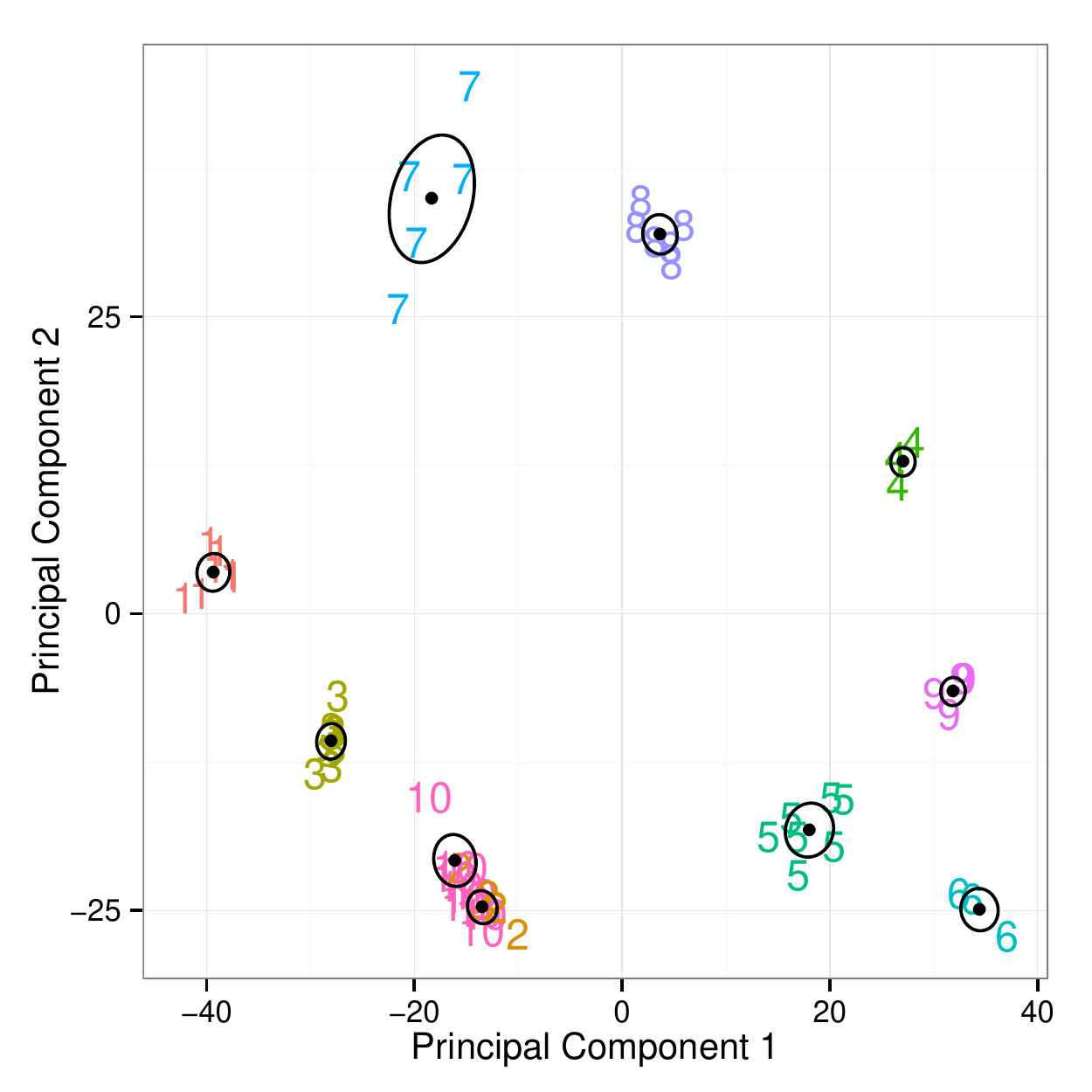}}
\subfloat[$\lambda = 10000$]{\label{fig:cremclust10000}
\includegraphics[scale=0.525]{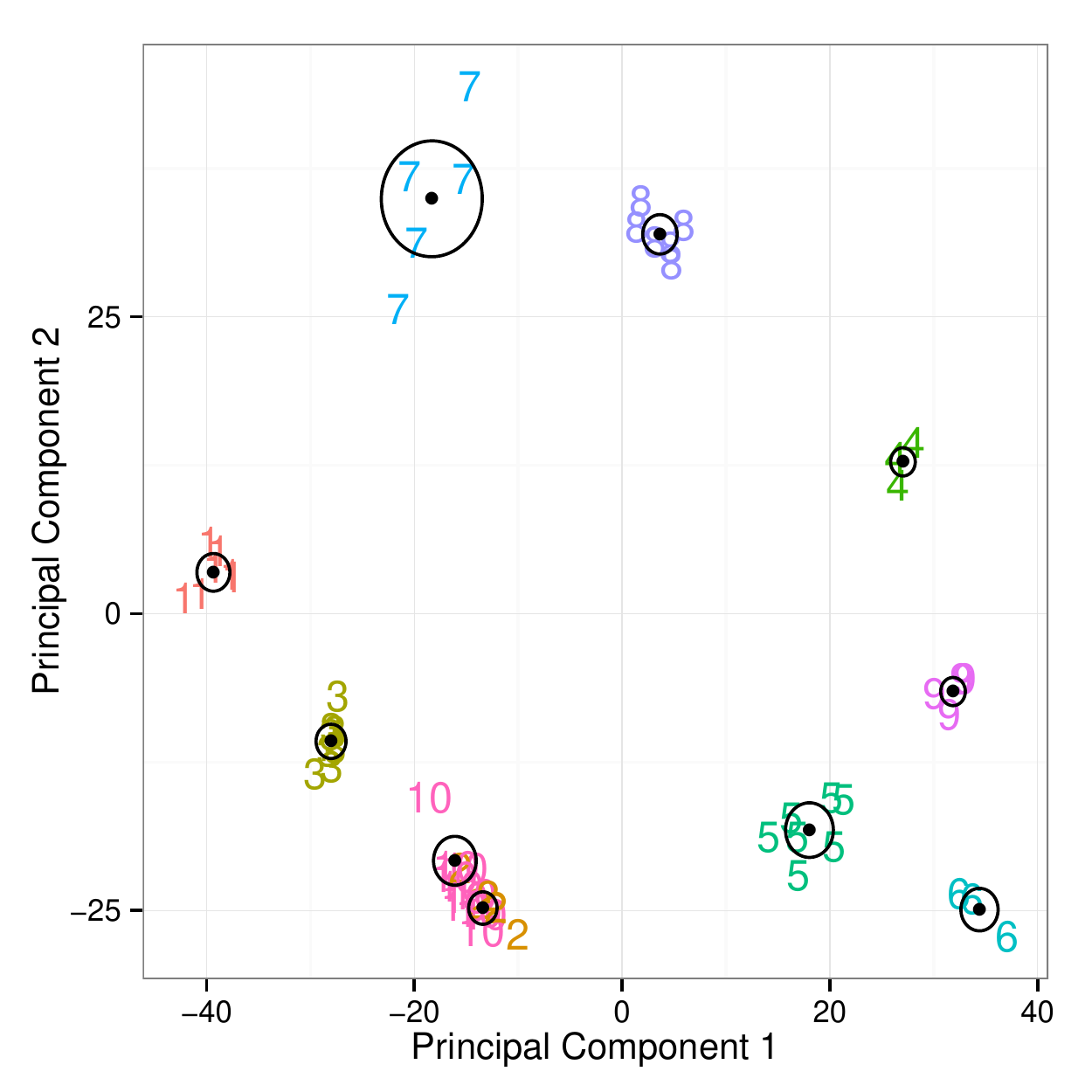}} \\
\end{tabular}
\caption{CERNN clustering projected onto the first two principal components of the data. Ellipses depict the first two eigenvectors (and their corresponding eigenvalues) of the estimated covariances of each cluster.}
\label{fig:clustering}
\end{figure}

Figure~\ref{fig:clustering} shows the results of clustering with our algorithm on a simulated data set. 
A total of 60 data points were generated from a mixture of 10 bivariate normals corresponding to 59 parameters in the most general case. The number of observations per cluster ranged from 3 to 11. We set the number of clusters $c$ to be $10$ and considered several different $\lambda$ values over a wide range (0.1, 10, 100, 10000). We could choose $\lambda$ in a completely data dependent way through cross-validation, but our main concern is to stabilize the procedure so fine-tuning $\lambda$ is not of paramount importance, especially when doing so complicates the procedure. We ran our algorithm 500 times using random initializations with the $k$-means++ algorithm \citep{ArtVas2007} and kept the clustering that gave the greatest value of the expected log likelihood 
\begin{eqnarray*}
\sum_{i=1}^n \sum_{k=1}^c w_{ik} \left [ \ln \pi_k + \ln \phi(\by_i \mid \bmu_k, \bSigma_k) \right ].
\end{eqnarray*}

The resulting clusterings are quite good over our broad range of $\lambda$ values. It is notable that for three out of the four values of $\lambda$, even clusters 2 and 10, which overlap, are somewhat distinguished. The only missteps occur when $\lambda = 10$, where cluster 1 is split into two clusters, and clusters 2 and 10 have been merged. The latter decision is reasonable given how much clusters 2 and 10 overlap.

\section{Discussion}
\label{sec:discussion}

The initial insight of \cite{Ste1975} has led to several methods for
shrinkage estimation of a sample covariance matrix $\bS$.  
These methods preserve the eigenvectors of $\bS$ while 
pushing $\bS$ towards a multiple of the identity matrix. Our Bayesian prior does precisely this
in a nonlinear fashion. Finite sample experiments comparing CERNN and CNR show that CERNN and CNR complement each other. CNR performs better when only a few eigenvalues of the population covariance are very large. CERNN performs better when there is a more even mix of high and low eigenvalues. Both perform at least as well and often better than the simple and asymptotically optimal LW estimator.

CERNN does require a singular value decomposition (SVD), as does CNR. Although highly optimized routines for accurately computing the SVD are readily available, such calculations are not cheap. Randomized linear algebra may provide computational relief \citep{HalMarTro2011,Mah2011}. If one can tolerate a small loss in accuracy, the SVD of a randomly sampled subset of the data or a random projection of the data can give an acceptable surrogate SVD.

Applications extend well beyond the classical statistical methods illustrated here. For example, in gene mapping with pedigree data, a covariance matrix is typically parameterized as a mixture of three components, one of which is the global kinship coefficient matrix capturing the relatedness between individuals in the study \citep{Lan2002}. The kinship matrix can be estimated from a high density SNP (single nucleotide polymorphism) panel rather than calculated from possibly faulty genealogical records. Because a typical study contains thousands of individuals typed at hundreds of thousands of genetic markers, this application occurs in the regime 
$n \ll p$. The construction of networks from gene co-expression data is another obvious genetic example \citep{Hor2011}. Readers working in other application areas can doubtless think of other pertinent examples. 


\section*{Acknowledgments}
This research was supported by NIH (United States Public Health Service)
grants GM53275 and HG006139.



\bibliographystyle{elsarticle-harv} 
\bibliography{CovarianceEstimation}

\begin{thebibliography}{43}
\expandafter\ifx\csname natexlab\endcsname\relax\def\natexlab#1{#1}\fi
\expandafter\ifx\csname url\endcsname\relax
  \def\url#1{\texttt{#1}}\fi
\expandafter\ifx\csname urlprefix\endcsname\relax\def\urlprefix{URL }\fi

\bibitem[{Arthur and Vassilvitskii(2007)}]{ArtVas2007}
Arthur, D., Vassilvitskii, S., 2007. $k$-means++: {T}he advantages of careful
  seeding. In: Proceedings of the eighteenth annual ACM-SIAM symposium on
  Discrete algorithms. SODA '07. Society for Industrial and Applied
  Mathematics, Philadelphia, PA, USA, pp. 1027--1035.

\bibitem[{Bache and Lichman(2013)}]{BacLic2013}
Bache, K., Lichman, M., 2013. {UCI} machine learning repository.
\newline\urlprefix\url{http://archive.ics.uci.edu/ml}

\bibitem[{Banerjee et~al.(2008)Banerjee, El~Ghaoui, and
  d'Aspremont}]{BanEl2008}
Banerjee, O., El~Ghaoui, L., d'Aspremont, A., Jun. 2008. Model selection
  through sparse maximum likelihood estimation for multivariate {G}aussian or
  binary data. Journal of Machine Learning Research 9, 485--516.

\bibitem[{Bickel and Levina(2008)}]{BicLev2008}
Bickel, P.~J., Levina, E., 2008. Regularized estimation of large covariance
  matrices. Annals of Statistics 36~(1), 199--227.

\bibitem[{Cai and Zhou(2012)}]{CaiZho2012}
Cai, T., Zhou, H., 2012. Minimax estimation of large covariance matrices under
  $\ell_1$ norm. Statistica Sinica 22, 1319--1378.

\bibitem[{Daniels and Kass(2001)}]{DanKas2001}
Daniels, M.~J., Kass, R.~E., 2001. Shrinkage estimators for covariance
  matrices. Biometrics 57~(4), 1173--1184.

\bibitem[{Dey and Srinivasan(1985)}]{DeySri1985}
Dey, D.~K., Srinivasan, C., 1985. Estimation of a covariance matrix under
  {S}tein's loss. Annals of Statistics 13~(4), 1581--1591.

\bibitem[{Fan et~al.(2011)Fan, Liao, and Mincheva}]{FanLiaMin2011}
Fan, J., Liao, Y., Mincheva, M., 2011. High-dimensional covariance matrix
  estimation in approximate factor models. Annals of Statistics 39~(6),
  3320--3356.

\bibitem[{Ferguson(1996)}]{Fer1996}
Ferguson, T.~S., 1996. A course in large sample theory. CRC Texts in
  Statistical Science Series. Chapman and Hall.

\bibitem[{Fraley and Raftery(2007)}]{FraRaf2007}
Fraley, C., Raftery, A.~E., Sep. 2007. Bayesian regularization for normal
  mixture estimation and model-based clustering. J. Classif. 24~(2), 155--181.

\bibitem[{Friedman et~al.(2008)Friedman, Hastie, and
  Tibshirani}]{FriHasTib2008}
Friedman, J., Hastie, T., Tibshirani, R., 2008. Sparse inverse covariance
  estimation with the graphical lasso. Biostatistics 9~(3), 432--441.

\bibitem[{Friedman(1989)}]{Fri1989}
Friedman, J.~H., 1989. Regularized discriminant analysis. Journal of the
  American Statistical Association 84~(405), 165--175.

\bibitem[{Guo et~al.(2007)Guo, Hastie, and Tibshirani}]{Guo2007}
Guo, Y., Hastie, T., Tibshirani, R., 2007. Regularized linear discriminant
  analysis and its application in microarrays. Biostatistics 8~(1), 86--100.

\bibitem[{Haff(1991)}]{Haf1991}
Haff, L.~R., 1991. The variational form of certain {B}ayes estimators. The
  Annals of Statistics 19~(3), 1163--1190.

\bibitem[{Halko et~al.(2011)Halko, Martinsson, and Tropp}]{HalMarTro2011}
Halko, N., Martinsson, P.~G., Tropp, J.~A., 2011. Finding structure with
  randomness: Probabilistic algorithms for constructing approximate matrix
  decompositions. SIAM Rev. 53~(2), 217--288.

\bibitem[{Hathaway(1985)}]{Hat1985}
Hathaway, R.~J., 1985. A constrained formulation of maximum-likelihood
  estimation for normal mixture distributions. Annals of Statistics 13~(2),
  795--800.

\bibitem[{Hero and Rajaratnam(2011)}]{HerRaj2011}
Hero, A., Rajaratnam, B., 2011. Large-scale correlation screening. Journal of
  the American Statistical Association 106~(496), 1540--1552.

\bibitem[{Hero and Rajaratnam(2012)}]{HerRaj2012}
Hero, A., Rajaratnam, B., 2012. Hub discovery in partial correlation graphs.
  Information Theory, IEEE Transactions on 58~(9), 6064--6078.

\bibitem[{Horvath(2011)}]{Hor2011}
Horvath, S., 2011. Weighted Network Analysis. Applications in Genomics and
  Systems Biology. Springer, New York.

\bibitem[{Huang et~al.(2006)Huang, Liu, Pourahmadi, and Liu}]{HuaLiuPou2006}
Huang, J.~Z., Liu, N., Pourahmadi, M., Liu, L., 2006. Covariance matrix
  selection and estimation via penalised normal likelihood. Biometrika 93~(1),
  85--98.

\bibitem[{Ingrassia(2004)}]{Ing2004}
Ingrassia, S., 2004. A likelihood-based constrained algorithm for multivariate
  normal mixture models. Statistical Methods and Applications 13~(2), 151--166.

\bibitem[{Ingrassia and Rocci(2007)}]{IngRoc2007}
Ingrassia, S., Rocci, R., 2007. Constrained monotone {EM} algorithms for finite
  mixture of multivariate {G}aussians. Computational Statistics \& Data
  Analysis 51~(11), 5339--5351.

\bibitem[{Khare and Rajaratnam(2011)}]{KhaRaj2011}
Khare, K., Rajaratnam, B., 2011. Wishart distributions for decomposable
  covariance graph models. Annals of Statistics 39, 514--555.

\bibitem[{Lange(2002)}]{Lan2002}
Lange, K., 2002. Mathematical and Statistical Methods for Genetic Analysis, 2nd
  Edition. Statistics for Biology and Health. Springer-Verlag, New York.

\bibitem[{Ledoit and Wolf(2004)}]{LedWol2004}
Ledoit, O., Wolf, M., 2004. A well-conditioned estimator for large-dimensional
  covariance matrices. Journal of Multivariate Analysis 88~(2), 365--411.

\bibitem[{Ledoit and Wolf(2012)}]{LedWol2012}
Ledoit, O., Wolf, M., 2012. Nonlinear shrinkage estimation of large-dimensional
  covariance matrices. Annals of Statistics 40~(2), 1024--1060.

\bibitem[{Mahoney(2011)}]{Mah2011}
Mahoney, M.~W., Feb. 2011. Randomized algorithms for matrices and data. Found.
  Trends Mach. Learn. 3~(2), 123--224.

\bibitem[{Mar\^{c}enko and Pastur(1967)}]{MarPas1967}
Mar\^{c}enko, V.~A., Pastur, L.~A., 1967. Distribution of eigenvalues for some
  sets of random matrices. Mathematics of the USSR-Sbornik 1~(4), 507--536.

\bibitem[{McLachlan and Peel(2000)}]{McL2000}
McLachlan, G., Peel, D., 2000. Finite Mixture Models. Wiley, New York.

\bibitem[{Mirsky(1975)}]{Mirsky1975}
Mirsky, L., 1975. A trace inequality of {John von Neumann}. Monatshefte f\"{u}r
  Mathematik 79, 303--306.

\bibitem[{Oh et~al.(2013)Oh, Rajaratnam, and Won}]{OhRaj2013}
Oh, S.-Y., Rajaratnam, B., Won, J.-H., 2013. CondReg: Condition Number
  Regularized Covariance Estimation. R package version 0.16.

\bibitem[{Peng et~al.(2009)Peng, Wang, Zhou, and Zhu}]{PenWanZho2009}
Peng, J., Wang, P., Zhou, N., Zhu, J., 2009. Partial correlation estimation by
  joint sparse regression models. Journal of the American Statistical
  Association 104~(486), 735--746.

\bibitem[{Pourahmadi(2011)}]{Pou2011}
Pourahmadi, M., 2011. Covariance estimation: The {GLM} and regularization
  perspectives. Statistical Science 26~(3), 369--387.

\bibitem[{Pourahmadi(2013)}]{Pou2013}
Pourahmadi, M., 2013. High-Dimensional Covariance Estimation: With
  High-Dimensional Data. Wiley.

\bibitem[{Pourahmadi et~al.(2007)Pourahmadi, Daniels, and Park}]{PouDanPar2007}
Pourahmadi, M., Daniels, M.~J., Park, T., 2007. Simultaneous modelling of the
  {C}holesky decomposition of several covariance matrices. Journal of
  Multivariate Analysis 98~(3), 568--587.

\bibitem[{Rajaratnam et~al.(2008)Rajaratnam, Massam, and
  Carvalho}]{RajMasCar2008}
Rajaratnam, B., Massam, H., Carvalho, C.~M., 2008. Flexible covariance
  estimation in graphical {G}aussian models. Annals of Statistics 36~(6),
  2818--2849.

\bibitem[{Ravikumar et~al.(2011)Ravikumar, Wainwright, Raskutti, and
  Yu}]{RavWaiRas2011}
Ravikumar, P., Wainwright, M.~J., Raskutti, G., Yu, B., 2011. High-dimensional
  covariance estimation by minimizing $\ell_1$-penalized log-determinant
  divergence. Electronic Journal of Statistics 5, 935--980.

\bibitem[{Rohde and Tsybakov(2011)}]{RohTsy2011}
Rohde, A., Tsybakov, A.~B., 2011. Estimation of high-dimensional low-rank
  matrices. Annals of Statistics 39~(2), 887--930.

\bibitem[{Sheena and Gupta(2003)}]{SheGup2003}
Sheena, Y., Gupta, A.~K., 2003. Estimation of the multivariate normal
  covariance matrix under some restrictions. Statistics \& Decisions 21~(4),
  327--342.

\bibitem[{Stein(1975)}]{Ste1975}
Stein, C., 1975. Estimation of a covariance matrix, 39th A. Meet. Institute of
  Mathematical Statistics, Atlanta.

\bibitem[{Warton(2008)}]{War2008}
Warton, D.~I., 2008. Penalized normal likelihood and ridge regularization of
  correlation and covariance matrices. Journal of the American Statistical
  Association 103~(481), 340--349.

\bibitem[{Won et~al.(2012)Won, Lim, Kim, and Rajaratnam}]{WonLimKim2012}
Won, J.-H., Lim, J., Kim, S.-J., Rajaratnam, B., 2012.
  Condition-number-regularized covariance estimation. Journal of the Royal
  Statistical Society: Series B (Statistical Methodology).

\bibitem[{Wu and Pourahmadi(2003)}]{WuPou2003}
Wu, W.~B., Pourahmadi, M., 2003. Nonparametric estimation of large covariance
  matrices of longitudinal data. Biometrika 90~(4), 831--844.

\end{thebibliography}

\end{document}